\documentclass[fleqn,10pt]{wlscirep}
\usepackage[utf8]{inputenc}
\usepackage[T1]{fontenc}
\usepackage{amsmath}
\title{The agnostic sampling transceiver}

\author[1,*]{Arijit Misra}
\author[1]{Janosch Meier}
\author[1]{Karanveer Singh}
\author[1]{Stefan Preu{\ss}ler}
\author[1]{Thomas Schneider}
\affil[1]{THz-Photonics Group, Technische Universit\"at Braunschweig, Schleinitzstra{\ss}e 22, 38106 Braunschweig, Germany}

\affil[*]{arijit.misra@ihf.tu-bs.de}

\begin{abstract}
Increasing capacity demands in the access networks require inventive concepts for the transmission and distribution of digital as well as analog signals over the same network. Here a new transceiver system, which is completely agnostic for the signals to be transmitted is presented. Nyquist sampling and time multiplexing of $N$ phase and intensity modulated digital and analog channels with one single modulator, as well as the transmission and demultiplexing with another modulator have been demonstrated. The aggregate symbol rate corresponds to the modulator bandwidth and can be further increased by a modification of the setup. No high-speed electronic signal processing or high bandwidth photonics is required. Apart from its simplicity and the possibility to process high bandwidth signals with low bandwidth electronics and photonics, the method has the potential to be easily integrated into any platform and thus, might be a solution for the increasing data rates in future access networks.    
\end{abstract}

\begin{document}

\flushbottom
\maketitle

\thispagestyle{empty}

\section*{Introduction}
According to the Cisco annual internet report\cite{cisco_2020}, 66\% of the 8 billion-world population in 2023 will use the internet and 77\% will do it by mobile devices. An unforeseeable worldwide crisis can push these growth rates even further and much faster. As an example, just in India Covid-19 has led to an increase in data traffic by 20\% in a few weeks. Online video streaming has increased by 120\% and gaming by more than 80\%. After the crisis has hit the country, internet usage, and data traffic in the internet exchange operator DE-CIX India has continually experienced new record levels with a shifting of the entire business traffic from the business districts into residential areas\cite{decix}. To fulfill such data demands and to improve the overall productivity and efficiency of the economy in normal circumstances, maximization of the bandwidth and efficiency of intra-datacenter communications\cite{Intrepid} and of the backbone and access networks is required. The intra-datacenter communication as well as the backbone and the access networks rely on digital optical networks\cite{Winzer:18}, whereas high capacity smart devices as well as broadband multimedia services have led to an unprecedented demand for high-speed wireless access to the networks\cite{RoF:Novak}. \par
Today, data transmission in the networks mainly relies on massive electronic digital signal processing. High sampling rate analog to digital converters (ADCs) in the transmitter and digital to analog converters (DACs) in the receiver, together with high bandwidth digital signal processors (DSP) are required for conventional optical links \cite{OIF}. Additionally, for the wireless access, the digital optical signal has to be transferred into the wireless domain and vice versa by high bandwidth electronics. The required bandwidth of the electronic signal processors and their energy consumption \cite{BaligaEC} increases with increasing data rates. In coherent fiber-optic communication systems, the bandwidth of the current DAC, for instance, puts a constraint on achieving much higher data rates\cite{DAC2020}.\par
Here we shall present a new method of signal transmission, which is completely independent on the signal and does not require high-speed electronic ADC, DAC, or DSP. As long as the Nyquist-Shannon sampling theorem limit\cite{book:oppenheim} is not violated and for ideal, noiseless components, each input signal in one channel will be multiplexed with other channels, transmitted and individually demultiplexed without any loss in information. The proposed agnostic transceiver is based on \textit{sinc}-pulse sequences. The modulation, sampling, and multiplexing of $N$ channels can be carried out with a single modulator. In this case, the symbol rate of all channels together corresponds to the bandwidth same as for orthogonal frequency division multiplexing (OFDM). For OFDM however, all channels in a superchannel have to be processed together, requiring high bandwidth digital signal processing in the transmitter and receiver. Here, just another single modulator driven with a radio frequency is needed for demultiplexing and the required receiver bandwidth is that of the single channels. If one separate modulator is used for pulse generation and another for sampling in each channel, the overall symbol rate can be 50\% higher than the bandwidth of the modulator. For digital signals the pulse repetition rate can be adapted to the symbol rate and therefore, even much higher baud rates can be achieved. However, in this case the transceiver would no longer be agnostic about the transmitted signal. Due to its simplicity and the possibility to achieve high data rates with low bandwidth electronics and photonics and without massive signal processing, full integration into silicon photonics platform is straightforward. \par

\begin{figure}[!htb] 
\centering
\includegraphics[width=\linewidth]{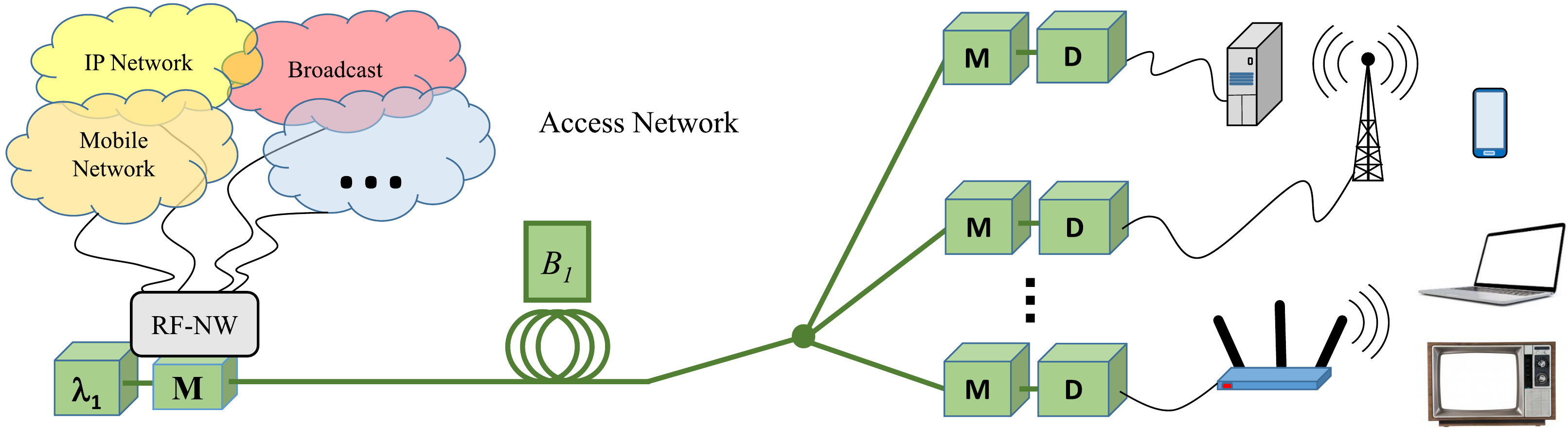}
\caption{Downlink in an access network with the proposed signal transmission. A number of $N$ digital, broadcast, wireless and other signals are multiplexed together with a radio frequency network (RF-NW) and one single modulator $M$ in the same rectangular bandwidth $B_1$ around the carrier wavelength $\lambda_1$. At the receiver side the single channel with the digital or analog signal is demultiplexed with a single modulator $M$ for each channel and converted back to the electrical domain with a detector $D$. Than the electrical signal is fed to the end device, a wireless access point or a broadcast antenna, where it might be amplified before transmission. The same might work for a possible uplink direction, around a wavelength $\lambda_2$ and the rectangular bandwidth $B_2$, which can be directly adjacent to $B_1$. The connection for the channels does not have to be pre-assigned. Each modulator at the right side receives all $N$ channels and can decide which one to demultiplex. Since the multiplexer and demultiplexer consists of a simple modulator, it can be integrated directly into a chip package. More connections and higher transmission data rates can be achieved by the incorporation of additional wavelengths. Due to the rectangular bandwidth of the channels, all of them can be multiplexed without any guard band between them.}
\label{fig:intro}
\end{figure}

The proposed transceiver is fully transparent for the transmitted signals. As long as the sampling theorem\cite{book:oppenheim} is not violated, the signal can be any form of a digital or analog signal, it can be a wireless signal for radio over fiber transmission or a digital one modulated in a higher-order modulation format. Since \textit{sinc}-pulse sequences are used, the spectrum of all multiplexed channels together is rectangular\cite{Soto2013, DaSilva2016}. Hence, in the same rectangular spectral band digital data and analog signals can be multiplexed together. To further enhance the transmittable data rate, these signals can additionally be multiplexed in the frequency domain without any guard band, using a microresonator frequency comb for instance\cite{Alishahi:19}. Like OFDM, the information can be transmitted with maximum possible symbol rate\cite{Nyquist1928}. However, no high speed DSP is required for the multiplexing or demultiplexing of the signals requiring DACs/ADCs operating at extremely high sampling rates. Thus, this proposed agnostic transceiver might provide a way to keep pace with the increasing demand and complexity of today’s communication networks. A possible scenario of the proposed signal transmission for the downlink in an access network can be seen in Fig.\,\ref{fig:intro}.  \par       

\section*{Principles}
\begin{figure}[!ht]
\centering
\includegraphics[width=\linewidth]{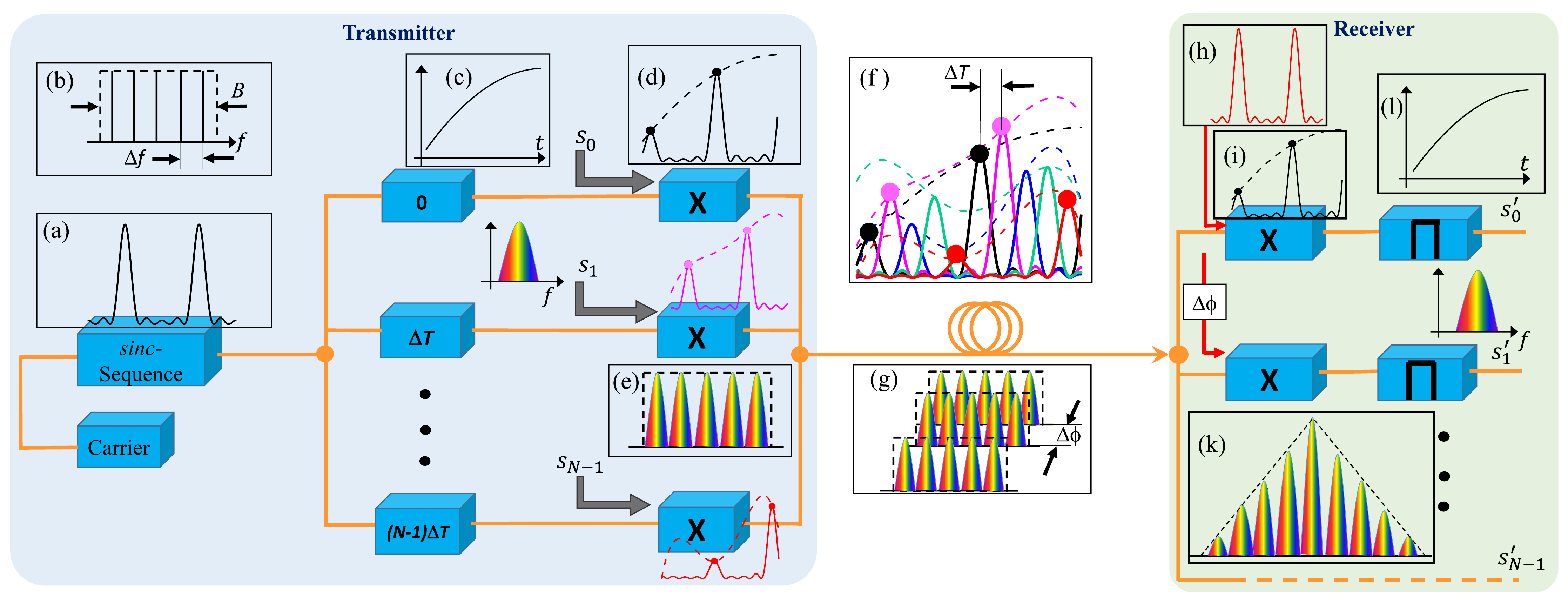}
\caption{Schematic illustration of the basic concept behind the proposed agnostic transceiver. In the transmitter (left side), an electrical, optical, microwave or THz carrier is modulated in a way, that a rectangular frequency comb with $N$ frequency lines and locked phases will be generated (b). In the corresponding time domain, this corresponds to a \textit{sinc}-pulse sequence (a). This sequence is divided into $N$ branches and delayed in time, so that all sequences are orthogonal to each other. In each branch the signal, representing the information to be transmitted ($s_0$ or (c)), will be multiplied with the \textit{sinc}-pulse sequence, resulting in a sampling of the signal ((d) in the time, and (e) in the frequency domain). All $N$ orthogonal channels will be multiplexed into the same rectangular spectral band (f) and (g), and transmitted. In the receiver (right side), all multiplexed channels (f) will be multiplied with a \textit{sinc}-pulse sequence with the correct time shift (h). Due to the orthogonality between all channels, the result is the retrieved individual sampled channel (i), or a triangular shaped spectrum resulting from the convolution of two rectangular spectra (k). Individual signals in their exact form are retrieved by filtering around the carrier with a band-pass filter of bandwidth same as the frequency spacing of the comb lines, or alternatively with a low-pass filter of half that bandwidth in the baseband after demodulation (l).}
\label{fig:principle}
\end{figure}

The proposed agnostic transceiver is based on the Nyquist-Shannon sampling theorem and orthogonality of \textit{sinc}-pulse sequences. Here we shall report the basic idea by using just one modulator for the whole transmitter. Nevertheless, to describe the basic principle, the schematic diagram in Fig. \ref{fig:principle} is more suitable. On the left hand side the transmitter (Tx) and on the right hand side the receiver (Rx) has been illustrated. A continuous wave (CW) carrier, which can be an electrical, optical, microwave, or THz wave, is modulated in a way that a \textit{sinc}-pulse sequence is generated. In the time domain, the \textit{sinc}-pulse sequence is an infinite summation of time shifted ideal \textit{sinc} pulses \cite{Soto2013, Soto2013a, Meier2019} (Fig. \ref{fig:principle}a). In the frequency domain, it corresponds to a rectangular frequency comb as shown in Fig. \ref{fig:principle}b. If the carrier is not suppressed, the comb always has an odd number of frequency lines. Such a comb with $N$ frequency lines and a frequency separation of $\Delta f$, has a total bandwidth of $ B= N \Delta f $. To generate such a comb it is sufficient to drive an intensity modulator with $n=\frac{N-1}{2}$ phase locked, equidistant sinusoidal frequencies \cite{Soto2013, Soto2013a}. The corresponding \textit{sinc}-pulse sequence in the time domain has a time duration of $ \Delta T = \frac{1}{B}$, from the peaks to the first zero crossings of the individual pulses, with a periodicity of $T = \frac{1}{\Delta f}$. Hence, there are $N-1$ zero crossings between two successive pulse peaks in the sequence. Analogous to ideal \textit{sinc} pulses, \textit{sinc}-pulse sequences are also orthogonal to each other, if the next sequence is time shifted by $\Delta T$ with respect to the previous one. Therefore, a comb with $N$ lines can be utilized for the transmission of information in $N$ independent, orthogonal channels occupying the same spectral width. Commonly, the \textit{sinc}-pulse sequence from a single source is divided into $N$ branches with proper time shifts to realize such orthogonally time interleaved systems \cite{Nakazawa2012, Hu:14, DaSilva2016}. But for the concept presented here, not the signal itself but its sampling values are multiplexed with proper time shifts and transmitted. \par 

As long as the signal is band-limited and the sampling theorem is not violated, the sampling values represent the original signal without any loss of information (amplitude and phase). The sampling process can be described in the frequency domain as the convolution of a frequency comb with the spectrum of the signal to sample ($s_1$ in Fig. \ref{fig:principle}) \cite{Preussler2016, Meier2019a, Meier2019}. Here the frequency comb consists of $N$ equidistant phase-locked lines with equal spectral power, and accordingly, the sampling generates $N$ equal copies of the input spectrum (Fig. \ref{fig:principle}e). In the time domain, the above process can be described as the multiplication between the signal (Fig. \ref{fig:principle}c) and a \textit{sinc}-pulse sequence (Fig. \ref{fig:principle}a), the result is the sampled signal (Fig. \ref{fig:principle}d) with the black dots representing the sampling values. Due to accurate time shifts, all sampling values are precisely in the zero crossings of all the other $N$ channels. Thus, the sampled signals are orthogonal to each other and can be multiplexed into the same frequency band, as shown in Fig. \ref{fig:principle}f and Fig. \ref{fig:principle}g in the time and the frequency domain, respectively. In the time domain, the $N$ different channels have a time shift of $\Delta T = \frac{T}{N}$ to each other, which corresponds to a phase shift of $\Delta \phi = \frac{2\pi}{N} $ between the channels.\par
\begin{figure}[!ht] 
\centering
\includegraphics[width=0.85\linewidth]{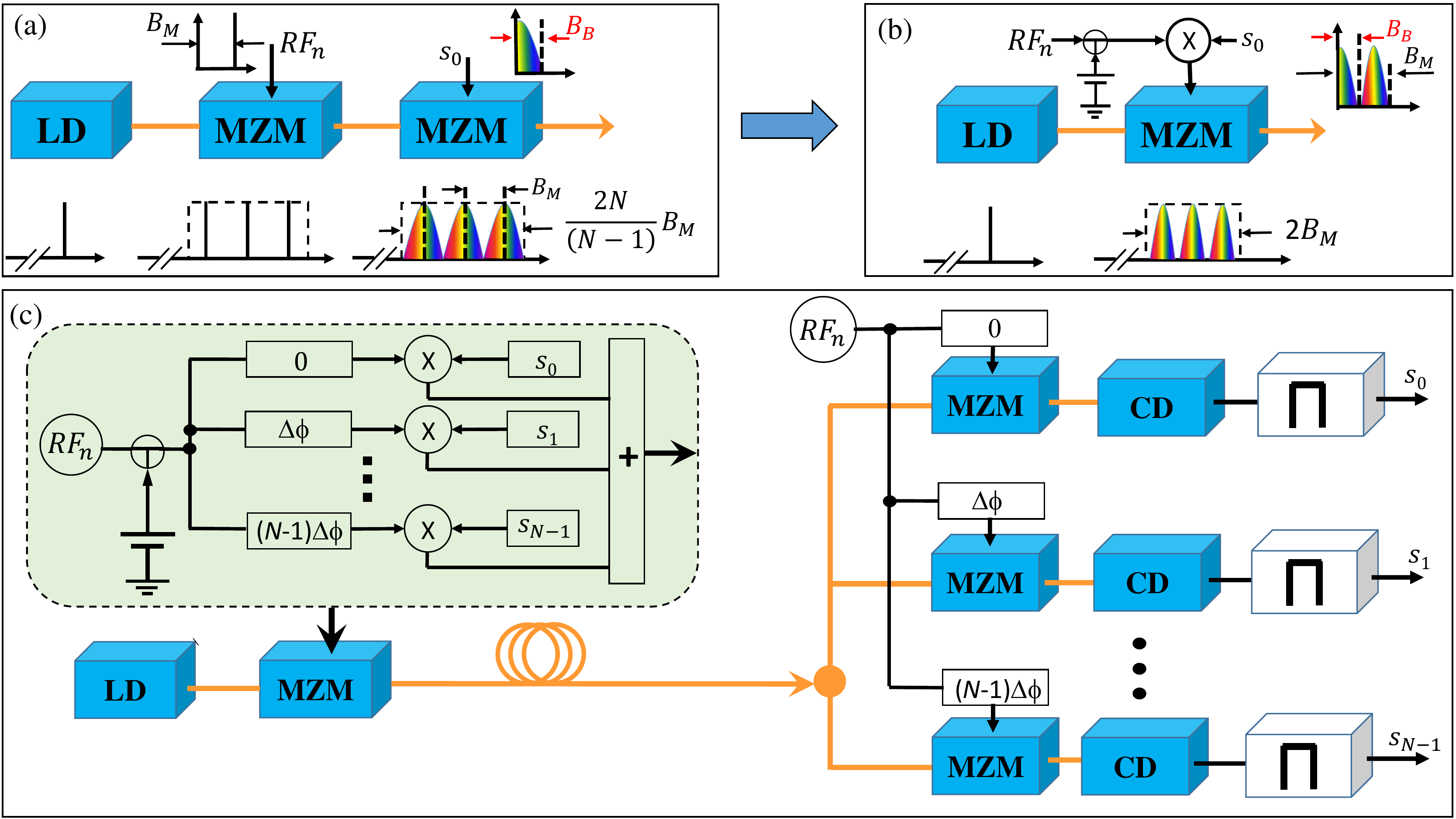}
\caption{Minimized setup for the agnostic sampling transceiver. The sinc pulse generation and sampling can be carried out in two consecutive (a) or one single modulator (b). The overall optical bandwidth in (a) is $2N/(N-1)$ times the maximum RF bandwidth of the incorporated modulators $B_M$ and therefore $3B_M$ for $N = 3$. For the single modulator the maximum bandwidth corresponds to $2B_M$. Accordingly, if three channels are multiplexed together, the overall symbol rate in these three channels together corresponds to $1.5B_M$ for (a) and $B_M$ for (b) (please see the methods section). The phase shifts of the electrical signals for all orthogonal channels have a difference of $\Delta \phi = 2\pi/N $ to each other. This phase shift ensures the correct time shift between the sinc pulse sequences. Thus, if all $N$ channels are multiplied with $n = (N-1)/2$ radio frequencies (RFs) with correct phase shifts, all the channels can be generated in the electrical domain and fed to one single modulator (c). Each channel is demultiplexed by a multiplication of all multiplexed channels with a sinc sequence with the correct time shift. This will be carried out in a single MZM driven with $n$ RF frequencies with the correct phase shift. A coherent detector and a filter in the baseband retrieves the original signal. MZM: Mach-Zehnder modulator, CD: coherent detector, LD: laser diode, $RF_n$ electrical source generating $n$ equidistant RF frequencies.}
\label{fig:principle2}
\end{figure}
In the receiver end (right hand side of Fig. \ref{fig:principle}) individual channels have to be demultiplexed and the original signal has to be retrieved from the sampling values. The multiplexed signal is divided into $N$ branches for simultaneous processing. In each branch, the corresponding channel will be demultiplexed by multiplying the multiplexed signal with a \textit{sinc}-pulse sequence \cite{Ecoc2015MUX}. Due to orthogonality, the channel, which has to be demultiplexed, is defined by the time shift of the sequence. Fig. \ref{fig:principle}i shows the demultiplexing of the first channel by multiplication with a \textit{sinc}-pulse sequence having correct bandwidth and time shift (Fig. \ref{fig:principle}h). In the equivalent frequency domain, this process corresponds to the convolution between the rectangular spectrum representing the $N$ copies of all $N$ channels (Fig. \ref{fig:principle}g) and another rectangular frequency comb. Hence, the result is a triangular shaped spectrum as depicted in Fig. \ref{fig:principle}k. In the last step, the signal spectrum around the carrier ($s_1^{\prime}$ in Fig. \ref{fig:principle}) has to be filtered out by a rectangular filter. If the filtering is carried out around the carrier frequency, the band-pass filter bandwidth should correspond to the repetition rate of the \textit{sinc}-pulse sequence ($\Delta f$). Alternatively, the filtering can take place in the baseband after demodulation, with a low-pass filter of bandwidth corresponding to half the repetition rate of the \textit{sinc}-pulse sequence ($\frac{\Delta f}{2}$).\par

If the sampling theorem \cite{book:oppenheim} is fulfilled and for ideal, noiseless components, the input signal in each channel (Fig. \ref{fig:principle}c) is equal to the output signal (Fig. \ref{fig:principle}l), please refer to the methods section for a complete mathematical description. Furthermore, the transceiver sets no precondition to the signal and the sampling method is also valid for amplitude and phase \cite{Meier2019}. Thus, any signal can be transmitted in any of the the $N$ channels. The signal can be an analog or a digital one; it can be amplitude or phase modulated, or premultiplexed in any format in the electrical domain. \par 
In this article, we shall show a proof of the concept with standard telecom equipment. However, as all single functionalities have already been shown in integrated silicon photonic chips \cite{Misra2019OE, Liu2020b}, the whole transceiver can be integrated easily on a silicon photonics or any other integrated photonics platform. \par
We have reduced the concept to the absolute minimum, so that just one single modulator undertakes all the functionalities of the $N$-branch transmitter and one single modulator in the receiver demultiplexes each channel. By an adjustment of the RF power and the bias voltage, intensity modulators, like Mach-Zehnder modulators (MZM) can be used to generate \textit{sinc}-pulse sequences with a very high quality \cite{Soto2013,Soto2013a}. A single modulator driven with $n$ equidistant sinusoidal radio frequencies generates a flat, phase-locked, rectangular frequency comb with $N=2n+1$ lines in the optical domain (Fig. \ref{fig:principle2}a). Thereafter, in a second MZM or I-Q modulator the pulse sequence can optically sample an electrical signal by a convolution between the comb and the signal spectrum. \par
Alternatively, as shown in Fig. \ref{fig:principle2}b, the signal can be multiplied with $n$ sinusoidal radio frequencies in an electrical mixer. With a proper DC level associated with the RF signals, the \textit{sinc}-pulse sequence generation as well as the sampling can be carried out together in one single MZM. For the signal in the next channel, all RF frequencies have a phase shift of $\Delta \phi = \frac{2\pi}{N} $. This phase shift can be adjusted by an electrical phase shifter. Therefore, all $N$ channels can be combined in the electrical domain as shown in Fig. \ref{fig:principle2}c, and one single intensity modulator is sufficient to transfer all channels into the optical domain. The electrical signal processing only requires an RF source, a bias-tee, electrical phase shifters, and mixers. The \textit{sinc}-pulse sequence generation, the time shifting, the sampling, and the multiplexing of all $N$ channels is carried out in one single modulator.\par 
In the receiver, every single channel can be demultiplexed by another MZM, driven with the RF frequencies with the right phase shifts. These frequencies have to be synchronized to the RF frequencies in the transmitter with a clock recovery. A coherent detector (CD) is utilized to transfer the signal from the optical to the electrical domain. This CD is required to detect the phase of the modulated signals. The sampled signal is transferred to the analog one either by a CD with sufficiently low bandwidth or by an additional filter in the electrical domain. 
\section*{Results}
\begin{figure}[!h]
\centering
\includegraphics[width=0.9\linewidth]{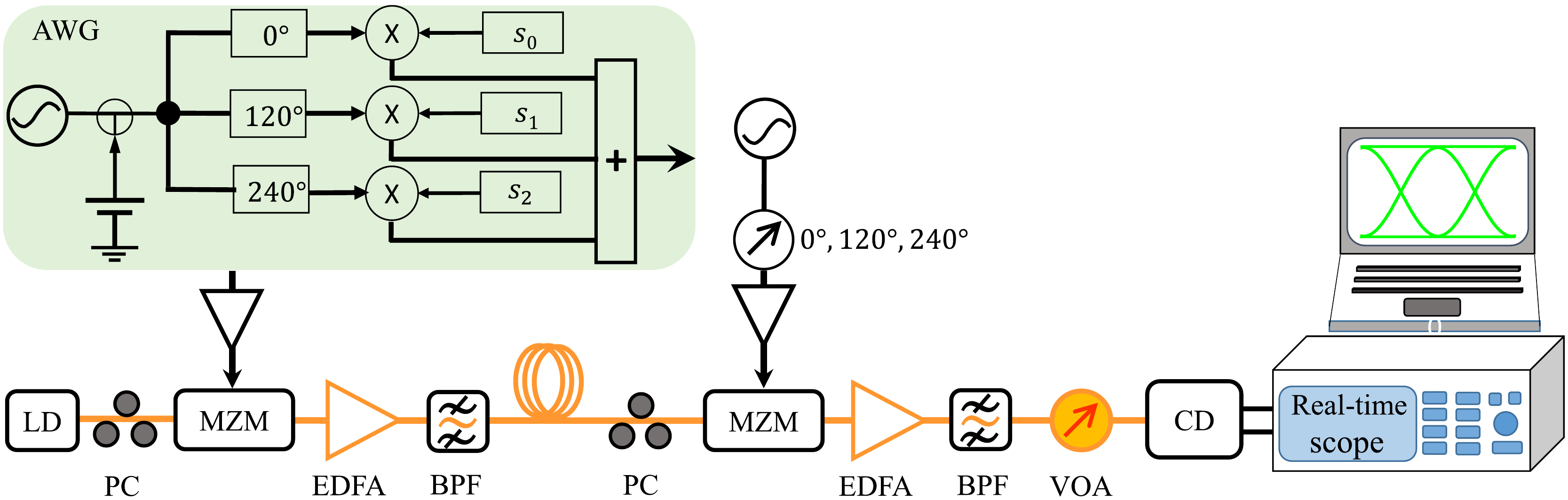}
\caption{Proof of concept experimental setup. Optically the transmitter consists of a laser source (LD) fixed at a wavelength of 1550.116\,nm and an MZM with a polarization controller (PC) while electrically it consists of an oscillator to generate a sinusoidal RF signal, a DC source and corresponding bias-tee, phase shifters for each channel and electrical mixers to mix the phase shifted RFs with the signal. However, for the case of experimental simplicity one arbitrary waveform generator (AWG 70000A, Tektronix) producing the three channel multiplexed signal has been used. This AWG limits the maximum achievable data rate for the proof of concept setup. A standard single mode fiber of different length was used as the transmission medium between the Tx and receiver (Rx). The receiver includes one MZM to demultiplex all the data channels one by one with proper phase shifts as described in Fig. \ref{fig:principle}h. The demultiplexed signal then goes through coherent detection and both the in-phase and quadrature components are recorded and stored in a real time oscilloscope for the signal classification carried out in a computer. Electrical amplifiers are used for necessary amplification of the electrical signals before injecting in to the modulators. Erbium doped fiber amplifiers (EDFA) along with 1\,nm band-pass filters (BPF) amplify the optical signal when required. The bit-error-rate (BER) performance against attenuation is investigated by a variable optical attenuator (VOA).}
\label{fig:setup}
\end{figure}
To experimentally demonstrate the underlying concept of the agnostic sampling transceiver we adopted an experimental set up (Fig. \ref{fig:setup}), analogous to the schematic in Fig. \ref{fig:principle2}c. Optically, the transmitter (Tx) section of the setup includes a laser diode (LD) and an MZM with proper polarization alignment. To avoid complexity, we did not implement polarization diversity. As described in the previous section, the $N$ number of channels can be sampled with \textit{sinc}-pulse sequences and multiplexed with one single modulator. To achieve this, the signal in the $N$ channels has to be multiplied with a phase-shifted RF signal and all the electrical signals have to be summed together. All of this can be accomplished with proper microwave circuitry and it can be integrated together with the modulator. In the proof of concept setup, however, we have used an arbitrary waveform generator (AWG) to generate the electrical signals and we have restricted the number of channels to $N=3$. Since the modulator generates $N=2n+1$ electrical lines, a single tone RF input $n=1$ is sufficient for the three channels. The signal generation by the AWG was restricted to that shown in Fig. \ref{fig:setup}. No electrical pre-compensation for the nonlinearities of different components or the chromatic dispersion was applied\cite{Misra2019OE}. \par
\begin{figure}[!htb]
\centering
\includegraphics[width=0.9\linewidth]{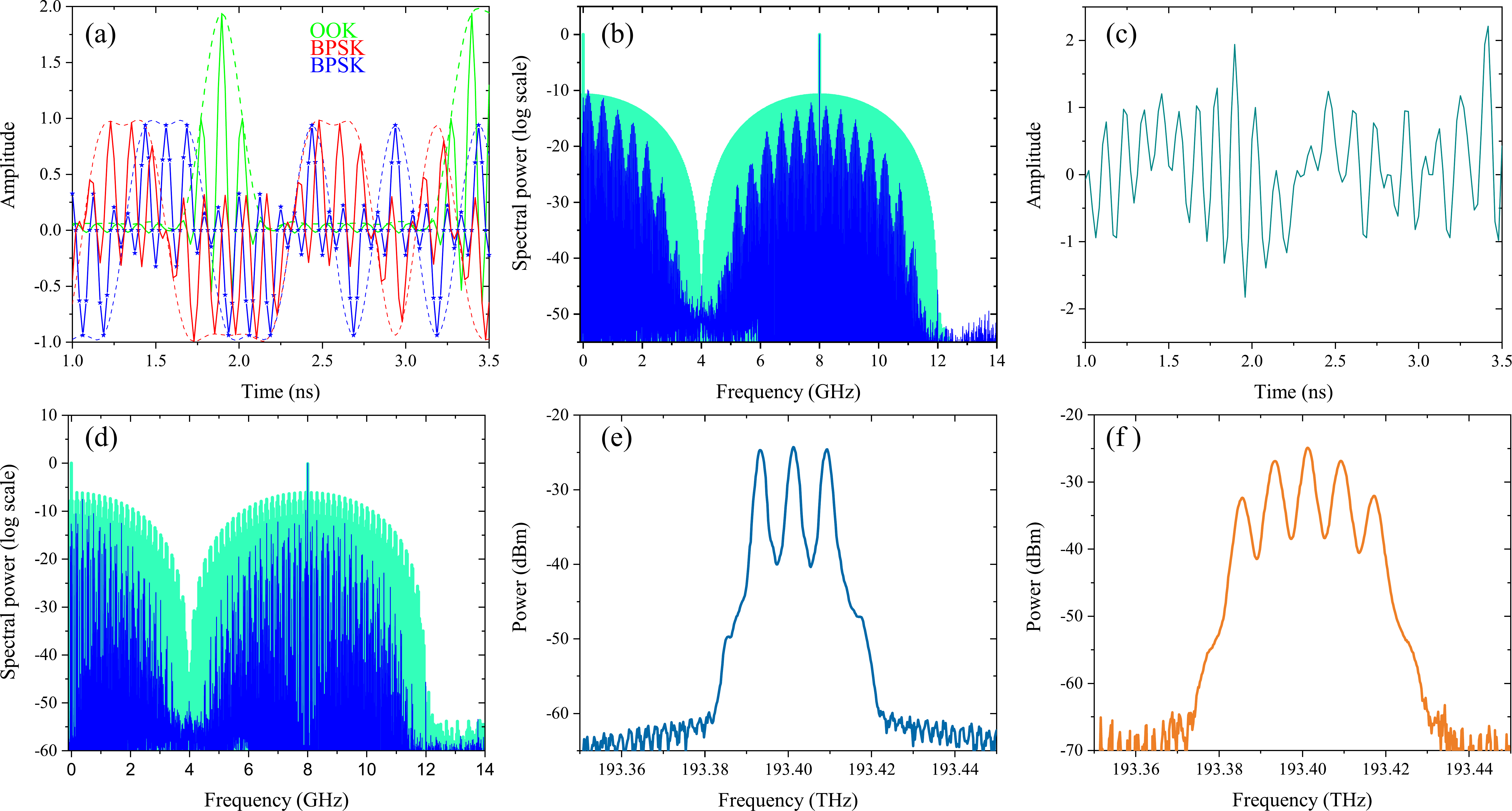}
\caption{The electrical data signals for each channel and the mixed signal in the RF domain are shown in dashed and solid curves respectively (a). The red and blue curves correspond to the 4\,Gbaud/s BPSK signals, while the green curve corresponds to the 4,\,Gbaud/s OOK signal. The base band spectrum of the RF mixed BPSK signal has been shown with the cyan curve representing the calculated and blue curve representing measured data at the output of the AWG (b).  After multiplexing two such BPSK and one OOK channels, a segment of the time domain signal is shown in (c), with the corresponding calculated and measured base band spectrum in (d). The rectangular optical spectrum of the three multiplexed channels sampled with the \textit{sinc}-pulse sequence is shown in (e).  Finally, the triangular shaped optical spectrum after demultiplexing one of the data channels is shown in (f).} 
\label{fig:signals}
\end{figure}
In the AWG we have generated three 4\,Gbaud pseudo-random bit sequences (PRBS-7), multiplied them with an 8\,GHz single tone sinusoidal RF signal with an additional DC level and multiplexed all the signals with a phase shift of $\Delta \phi=120^\circ$ with respect to each other. A segment of the generated electrical signal for one channel with on-off keying (OOK, green) and two channels with binary phase shift keying (BPSK, red and blue) modulated pseudo-random bit sequences (PRBS-7) can be seen in Fig. \ref{fig:signals}a. The respective data signals have been shown in dashed curves. As can be seen, the sinusoids (solid curves in Fig. \ref{fig:signals}a) are having proper DC thresholds and they are time shifted to each other by one-third of the period. As we were limited by the sampling rate of 48\,GS/s by the AWG, the stars of the blue curve show 6 points per period, indicating the 8\,GHz sinusoidal signal. The cyan curve in Fig. \ref{fig:signals}b corresponds to the fast Fourier transform of the time domain signal for the BPSK that was loaded to the AWG. The corresponding output signal from the AWG exhibited spectrum is shown in blue. All the three channels together, with the underlying $8$\,GHz sinusoidal frequency, can be seen in Fig. \ref{fig:signals}c with the base band spectrum as shown in Fig. \ref{fig:signals}d. When transferred to the optical domain by an MZM with the right bias and RF power, the generated three-line spectrum is flat and fits in a rectangular bandwidth (see Fig. \ref{fig:signals}e). As described in Fig. \ref{fig:principle}g, the optical spectrum contains the information of all three channels in the same rectangular bandwidth with a phase shift of $\Delta \phi=120^\circ$ between the channels. For the demultiplexing of one of the channels in the receiver, the multiplexed channels will be multiplied with a frequency comb with the proper phase shift. This is done in the receiver MZM by driving it with an 8\,GHz sinusoidal RF signal, which was synchronized to the transmitter. This synchronisation can easily be achieved by an efficient clock recovery mechanism. As described in Fig. \ref{fig:principle2}k, the multiplication in the time domain corresponds to the convolution between two rectangular functions in the frequency domain, which results in a triangular spectrum, as shown in Fig. \ref{fig:signals}f. \par 

\begin{figure}[!htb]
\centering
\includegraphics[width=0.85\linewidth]{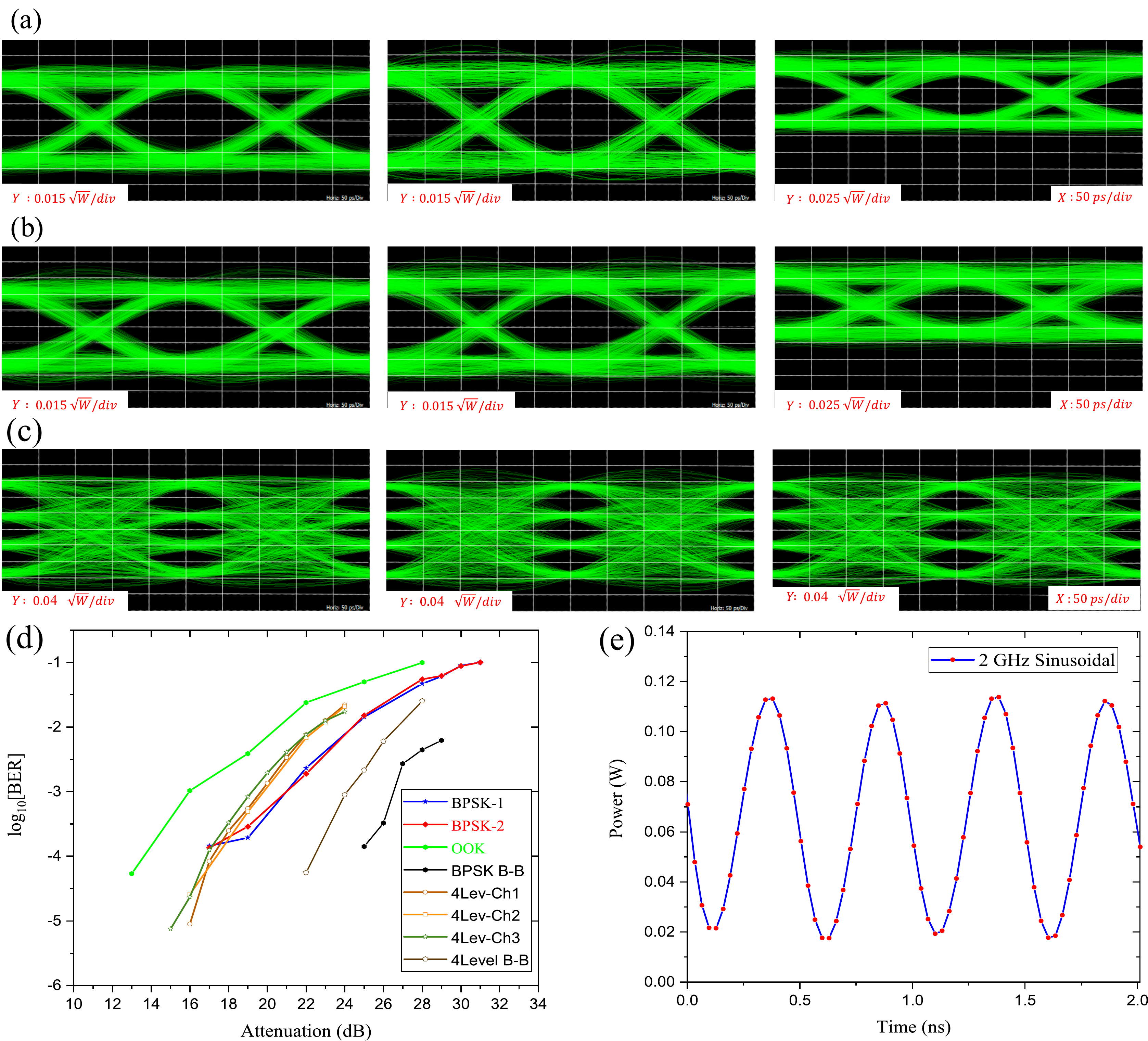}
\caption{ Eye diagrams of the received \textit{sinc}-pulse sequence sampled 4\,Gbaud/s optical BPSK (left and middle column) and OOK signals (right) after  5\,km (a) and 10\,km (b) of SMF transmission and demultiplexing with a phase matched \textit{sinc}-pulse sequence. In (c) the eye diagrams of the received demultiplexed channels for three channels modulated with a two level BPSK after 5\,km fiber transmission are shown. No digital chromatic dispersion compensation was needed for `zero' BER in more than 300,000 measured bits. The BER performance against attenuation of the received \textit{sinc}-pulse sequence sampled signals after demultiplexing shows a better performance of the BPSK signal compared to the OOK signal, due to limited Euclidean distance (d). Along with digital phase and intensity encoded data in two channels a third channel with an analog 2\,GHz sinusoidal signal has been transmitted successfully (e).} 
\label{fig:eyes}
\end{figure}
In the first experiment, we have modulated all three channels with a 4\,Gbaud/s, pseudo-random bit sequence (PRBS-7). The first two channels were modulated in a binary phase shift keying (BPSK) and the third in an on-off keying (OOK) format. The eye diagrams for the two BPSK (left and middle) and the OOK modulation (right) as received and demultiplexed successively after 5\,km and 10\,km fiber transmission are depicted in Fig. \ref{fig:eyes}a and \ref{fig:eyes}b, respectively. For the different fiber lengths, the input optical power at the receiver was maintained to be the same.  The error vector magnitudes (EVMs) for the BPSK signals were measured to be 9.66\% and 12.98\% after 5\,km and 10\,km of standard single mode fiber (SMF) transmission, respectively. These EVM values ensure an error free transmission \cite{SchmogrowEVM}, which was consistent with the BER measurements.  It is worth mentioning, that for all the measurements a digital compensation of chromatic dispersion, as well as MZM nonlinearity, was not applied. The BER performance against the attenuation at the VOA (see Fig. \ref{fig:setup}) for all three channels and a fiber length of 5\,km has been presented in Fig. \ref{fig:eyes}d, by taking the 4\,dB insertion loss of the VOA into account. As expected, the two BPSK channels (BPSK-1 and -2) show almost the same performance. Due to the lower Euclidian distance of the symbols in the modulation, the OOK modulation has a slightly lower performance. \par
In the next experiment, all three channels were modulated with a two-amplitude 4\,Gbaud/s, PRBS-7 BPSK signal, resulting in a four-level modulation. The demultiplexed channels after 5\,km fiber transmission can be seen in Fig. \ref{fig:eyes}c. For the four-level data signal, the average EVM was around 4.5\% after 5\,km of transmission. Again, we have measured no errors in 300,000 transmitted Bit. The BER performance depending on the attenuation can be seen in Fig. \ref{fig:eyes}d. As expected, all three channels show an almost identical behavior (4Lev-Ch1, 2, 3). \par 
The agnostic transceiver is capable of transmitting any type of signal as long as the signal satisfies the Nyquist criterion. Thus, in the last experiment, we have transmitted a 4\,Gbaud/s, PRBS-7 BPSK signal in the first channel (phase modulation), a 4\,Gbaud/s, PRBS-7 OOK signal in the second (intensity modulation) and an analog 2\,GHz sinusoidal signal in the third channel. The resultant retrieved signal for the analog channel is shown in Fig. \ref{fig:eyes}e. The two digital data channels associated with it did not show any bit error. 

\section*{Discussion}
We have presented a new kind of transceiver, which is completely agnostic for the signal to be transmitted. In just one single modulator the signals of $N$ different electrical channels, which can be intensity or phase modulated digital or analog ones, are optically sampled and multiplexed with \textit{sinc}-pulse sequences. Each channel can be demultiplexed with another modulator at the receiver end. No optical delay lines or phase shifters and no broadband digital signal processing is required. Due to the simplicity of the method, integration into silicon photonics or any other platform is straight-forward. \par 
For the sake of avoiding complexity and due to a lack of experimental equipment, we have presented the concept just for an intensity modulator. For higher-order modulation formats, like quadrature amplitude modulation (QAM), the single modulator in the transmitter has to be replaced by an in-phase, quadrature-phase (I-Q) modulator and an additional electrical network, similar to that shown in the green box of Fig. \ref{fig:setup}, is necessary for the Q-branch of the modulator. For the Q-branch of the electrical network, however, the same sinusoidal signal and the same phase shift is required. Accordingly, just an additional electrical mixing step is necessary. Please note, that I and Q are defined by a $90^{\circ}$ phase difference of the carrier, whereas the time shift of the \textit{sinc}-pulse sequences is dependent on the phase of the sinusoidal waves used for their generation. Thus, for the demultiplexing of I-Q modulated signals a single multiplication with the \textit{sinc}-pulse sequence with the correct time shift in a single intensity modulator is sufficient.\par
Carrying out all functionalities in one single modulator has the advantage of being extremely simple. However, the maximum symbol rate that can be transmitted in all $N$ channels together corresponds to the RF bandwidth of the modulator (please see the methods section). A symbol rate of up to 1.5 times the RF bandwidth of the modulator is possible if the \textit{sinc}-pulse generation and sampling are carried out in two different modulators. Therefore, with one integrated modulator with an RF bandwidth of 500 GHz \cite{500GHZMZM}, $N$ multiplexed signals with an overall symbol rate of 0.5 Tbaud/s can be multiplexed and transmitted. With a modulator for pulse generation and one for sampling, the symbol rate could increase to 0.75 Tbaud/s. By adapting the signal to the sampling pulses, for digital signals, a much higher symbol rate is possible for the single and the cascaded modulators. However, due to the violation of the sampling theorem, the advantages of being an agnostic transceiver have to be sacrificed \cite{Preussler_Nyphe}. 
For a polarization multiplexing, the setup can easily be doubled for the second polarization. All $N$ multiplexed channels are in the same rectangular bandwidth and a wavelength division multiplexing of these channels without any guard band is possible \cite{DaSilva2016}. Hence, the method might lead to a new way of signal transmission, especially in access networks were digital and analog signals are transmitted together, to keep pace with the increasing data rate demands of tomorrow. \par
\section*{Methods}
\subsection*{Transmittable symbol rate}
If the \textit{sinc}-pulse generation and multiplexing of all $N$ channels are carried out with just one modulator with an RF bandwidth of $B_M$, as described above, the maximum bandwidth of the generated optical frequency comb convoluted with the signal spectrum corresponds to $2B_M$, as shown in Fig. \ref{fig:principle2}b. In this case, the maximum baseband bandwidth of the signal in the single channel, or the symbol rate can be $B_B=\frac{B_M}{N}$. Thus, the combined symbol rate of all $N$ channels together corresponds to the bandwidth of the modulator $B_M$. \par
The symbol rate can be increased if the pulse generation and sampling are carried out in two consecutive modulators as shown in Fig. \ref{fig:principle2}a. If one single modulator is driven with $n$ equidistant sinusoidal radio frequencies for pulse generation, the modulator generates a rectangular frequency comb with $N=2n+1$ lines in the optical domain. As depicted in Fig. \ref{fig:principle2}a, the maximum bandwidth of the comb convoluted with the signal spectrum corresponds to $\frac{2N}{N-1}\times B_M$, with $B_M$ as its maximum RF bandwidth, or $3 B_M$ for $N = 3$ and $n = 1$. The maximum baseband bandwidth of the signal in the single channel can be $B_B=\frac{1}{N-1} B_M$. Thus, for $N = 3$, the symbol rate for all three channels together would be $1.5 \times B_M$ and therefore $50\%$ higher than the RF bandwidth of the single modulator.

\subsection*{Theory of the agnostic sampling transceiver}
In the agnostic transmitter, the signals of $N$ channels will be sampled with \textit{sinc}-pulse sequences and, by exploiting the orthogonality of the \textit{sinc}-pulse sequences, they will be multiplexed by time interleaving into the same rectangular spectral band. In the receiver, again by exploiting the orthogonality of the \textit{sinc}-pulse sequences, the signal of the single channel will be demultiplexed and the original signal will be retrieved from the sampling values. Here we will mathematically prove that for ideal components this process can be carried out without any loss in information.\par
According to the sampling theorem, each bandwidth-limited signal can be represented by an infinite summation of \textit{sinc} pulses weighted with the sampling points:
\begin{equation}
\label{sampling theorem}
f(t)=\sum_{k=-\infty}^\infty f\left(\frac{k}{\Delta f_\mathrm{s}}\right)\cdot\mathrm{sinc}(\Delta f_\mathrm{s}t-k)\ .
\end{equation}
The signal has to be limited in frequency by a baseband width of $\frac{\Delta f_\mathrm{s}}{2}$ and for avoiding a discontinuity at $t=0$ the \textit{sinc} function can be defined as: 
\begin{equation}
\label{sinc def}
\mathrm{sinc}(t)=\lim_{\substack{x\to t\\x\in\mathbb{R},x\neq t}}\left(\frac{\mathrm{sin}(\pi x)}{\pi x}\right)\ .
\end{equation}
Thus, without any loss in information, a signal with the corresponding bandwidth is completely represented by its sampling points and it is sufficient to proof, that the sampling points are preserved.\par
The sampling theorem is based on ideal \textit{sinc} pulses, which are infinitely stretched in the time domain and hence just a mathematical construct. Here the signal is sampled and multiplexed with \textit{sinc}-pulse sequences instead. However, \textit{sinc}-pulse sequences can be seen as an infinite superposition of ideal \textit{sinc} pulses \cite{Soto2013}:
\begin{equation}
\label{sq def}
\mathrm{sq}_{N,B}(t)=\sum_{k=-\infty}^\infty \mathrm{sinc}(Bt-kN)=\frac{2}{N}\left(\frac{1}{2}+\sum_{k=1}^\frac{N-1}{2}\mathrm{cos}\left(\frac{2\pi kBt}{N}\right)\right)\ ,
\end{equation}
where $B$ is the bandwidth of the \textit{sinc}-pulse sequence and $N$ an odd number of comb lines. The periodicity of the \textit{sinc}-pulse sequence is given by $T=\frac{N}{B}$, while \mbox{$\Delta T=\frac{1}{B}$} defines the time between a peak and the first zero crossing. The Fourier transform of the \textit{sinc}-pulse sequence is a rectangular frequency comb with an odd number of frequency lines $N$ given by:
\begin{equation}
\label{Comb def}
\mathrm{Comb}_{N,B}(f)=[\mathcal{F}_t(\mathrm{sq}_{N,B}(t))](f)=\frac{1}{N}\sum_{k=-\frac{N-1}{2}}^\frac{N-1}{2}\delta\left(f-k\frac{B}{N}\right)
\end{equation}
with $\delta$ as the Dirac delta function and $\mathcal{F}_t$ as the Fourier transform operator.\par Ideal \textit{sinc} pulses and \textit{sinc}-pulse sequences have in common that they are orthogonal to each other. Additionally, in case of an ideal \textit{sinc} pulse, the sampling value is defined by its peak, whereas for a \textit{sinc}-pulse sequence the sampling values are defined by the peaks of the pulses along sequence \cite{Meier2019}. The main difference however is, that an infinite number of ideal \textit{sinc} pulses can be multiplexed, whereas the number of multiplexable \textit{sinc}-pulse sequences is limited to $N$.\par
Without any crosstalk between the sampling points, multiplexing of $N$ channels $s_0,...,s_{N-1}$, sampled with \textit{sinc}-pulse sequences with the time shift of $\Delta T$ relative to each other, can be formulated as:
\begin{equation}
\label{multiplexed signal}
s(t)=\sum_{k=0}^{N-1}s_{k}(t)\cdot\mathrm{sq}_{N,B}(t-k\Delta T)\ .
\end{equation}
Here baseband bandwidth limitation of each channel is $\frac{B}{2N}$. All $N$ multiplexed channels together $s(t)$ have a baseband bandwidth of $B/2$, which corresponds to half the bandwidth of the \textit{sinc}-pulse sequences.  \par
\begin{figure}[!htb]
\centering\includegraphics[width=0.75\textwidth]{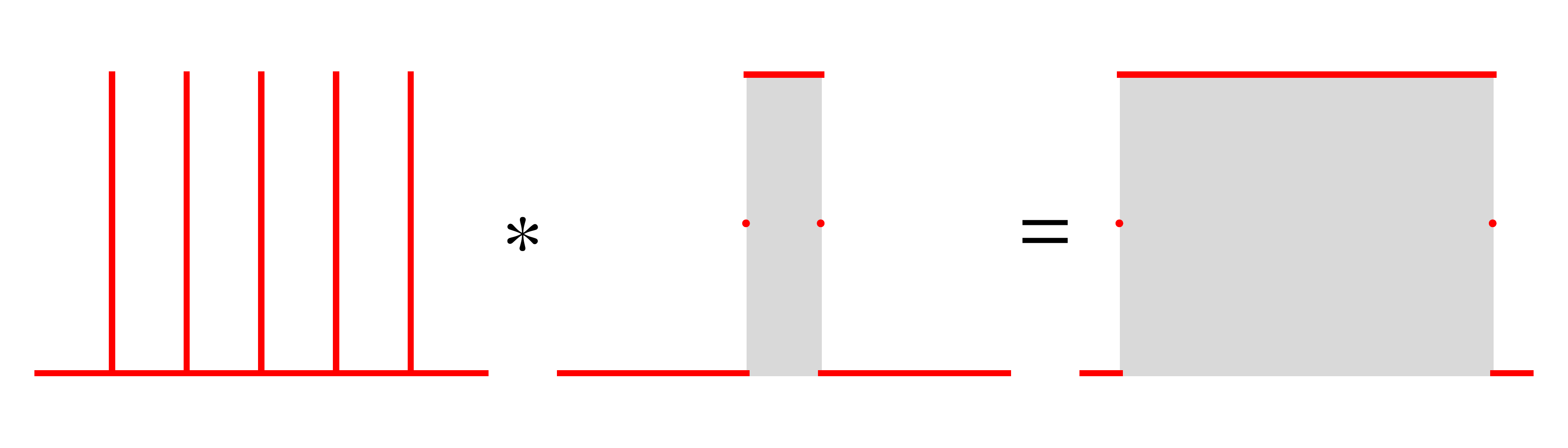}
\caption{Frequency domain representation of the multiplication between a \textit{sinc}-pulse sequence and a \textit{sinc} pulse as in Eq. (\ref{multiplication theorem}). The finite frequency comb convoluted with a rectangular function with the same bandwidth as the frequency spacing between the comb lines, results in a wider rectangular function.}
\label{fig:method}
\end{figure}

In the receiver single channel with a time shift of $l\Delta T$ with $l\in{0,...,N-1}$ will be extracted by a multiplication of all multiplexed channels $s(t)$ with a \textit{sinc}-pulse sequence with the same time shift. In an additional step, the signal will be filtered by a rectangular filter with a bandwidth of $\frac{B}{2N}$. By using the convolution theorem, the demultiplexed signal can be written as:
\begin{equation}\label{output}
\begin{aligned}
x_l(t)&=\left[\mathcal{F}_f^{-1}\left(\left[\mathcal{F}_t(s(t)\cdot\mathrm{sq}_{N,B}(t-l\Delta T))\right](f)\cdot\Pi_{B/N}(f)\right)\right](t)
\\
&=\left[\left[\mathcal{F}_f^{-1}\left(\left[\mathcal{F}_t(s(t)\cdot\mathrm{sq}_{N,B}(t-l\Delta T))\right](f)\right)\right]*\left[\mathcal{F}_f^{-1}\left(\Pi_{B/N}(f)\right)\right]\right](t)\ .
\end{aligned}
\end{equation}
where $\Pi_b(f)$, with $b=B/N$, designates the rectangular function equal to 1 for $|f|<\frac{1}{2}b$, \mbox{$\frac{1}{2}$ for $|f|=\frac{1}{2}b$} and zero elsewhere and $\mathcal{F}_f^{-1}$ is the inverse Fourier transform operator. Following Eq. (\ref{output}), one sampling point $x_l(t_\mathrm{s})$ at a sampling instance \mbox {$t_\mathrm{s}=l\Delta T+mN\Delta T$} with an integer $m$, is given as:
\begin{equation}\label{output at sampling points 1} 
\begin{aligned}
& x_l(t_\mathrm{s})=\int_{-\infty}^\infty s\left(t_\mathrm{s}-\tau\right)
\cdot\mathrm{sq}_{N,B}(t_\mathrm{s}-\tau-l\Delta T)\cdot\frac{B}{N}\mathrm{sinc}\left(\frac{B\tau}{N}\right)\mathrm{d}\tau\ .
\\
\end{aligned}
\end{equation}
Considering periodicity and symmetry property of the \textit{sinc}-pulse sequence it follows:
\begin{equation}\label{output at sampling points 2}
\begin{aligned}
& x_l(t_\mathrm{s})=\int_{-\infty}^\infty s(t_\mathrm{s}-\tau)\cdot\mathrm{sq}_{N,B}(\tau)\cdot\frac{B}{N}\mathrm{sinc}\left(\frac{B\tau}{N}\right)\mathrm{d}\tau\ .
\\
\end{aligned}
\end{equation}
According to the multiplication theorem between \textit{sinc}-pulse sequences and \textit{sinc} pulses conceptually depicted in Fig. \ref{fig:method} in the frequency domain, the expression above can be rewritten as:
\begin{equation}\label{output at sampling points 3}
x_l(t_\mathrm{s})=\int_{-\infty}^\infty s(t_\mathrm{s}-\tau)\cdot\frac{B}{N}\mathrm{sinc}(B\tau)\ \mathrm{d}\tau\ .
\end{equation}
As discussed earlier, all channels together $s(t)$ are baseband bandwidth-limited by $\frac{B}{2}$, thus, using the ideal sampling with \textit{sinc} pulses, it follows:
\begin{equation}\label{output at sampling points 4}
x_l(t_\mathrm{s})=\frac{1}{N}s(t_\mathrm{s})\ .
\end{equation}
Due to the orthogonality of the \textit{sinc}-pulse sequences, this is identical to:
\begin{equation}\label{output at sampling points 5}
x_l(t_\mathrm{s})=\frac{1}{N}s_l(t_\mathrm{s})\ .
\end{equation}
Therefore, for the peak positions or sampling instances \mbox {$t_\mathrm{s}=l\Delta T+mN\Delta T$} the sampling points are preserved. Except for a multiplication with the constant factor $\frac{1}{N}$ they are invariant. Here for the derivation of Eq. (\ref{output at sampling points 3}) a multiplication theorem between \textit{sinc}-pulse sequences and \textit{sinc} pulses has been used. This multiplication theorem can be understood in the frequency domain. Since a Dirac delta function inside a convolution acts like an identity operator, it follows:
\begin{equation}\label{multiplication theorem}
\begin{aligned}
\mathrm{sq}_{N,B}(\tau)\cdot\frac{B}{N}\mathrm{sinc}\left(\frac{B\tau}{N}\right)
&=\left[\mathcal{F}_f^{-1}\left(\left[\mathrm{Comb}_{N,B}*\Pi_{B/N}\right](f)\right)\right](\tau)
\\
&=\left[\mathcal{F}_f^{-1}\left(\frac{1}{N}\Pi_B\left(f\right)\right)\right](\tau)
\\
&=\frac{B}{N}\mathrm{sinc}(B\tau).
\end{aligned}
\end{equation}
The underlying idea behind this multiplication theorem is visualized in Fig. \ref{fig:method} in the frequency domain.
According to Eq. (\ref{output at sampling points 5}), except for the factor $\frac{1}{N}$, the retrieved signal at the receiver $x_l(t_\mathrm{s})$ after sampling, multiplexing and demultiplexing with \textit{sinc}-pulse sequences is identical to the input signal at sampling instances $s_l(t_\mathrm{s})$. If the single input signal $s_l(t)$ is sampled and multiplexed with ideal \textit{sinc} pulses, it follows:
\begin{equation}\label{sampling theorem for a single channel}
\begin{aligned}
s_l(t)&=\sum_{k=-\infty}^\infty s_l\left(\frac{kN}{B}+l\Delta T\right)\cdot\mathrm{sinc}\left(\frac{B}{N}(t-l\Delta T)-k\right)
\\
&=\sum_{k=-\infty}^\infty s_l(kN\Delta T+l\Delta T)\cdot\mathrm{sinc}\left(\frac{1}{N\Delta T}(t-l\Delta T)-k\right)\ .
\end{aligned}
\end{equation}
Therefore, with $j$ as an integer, $s_l(t)$ is completely defined by the sampling points \mbox{$s_l(jN\Delta T+l\Delta T)$} of the \textit{sinc}-pulse sequence used for multiplexing.\par
Due to the filter, the demultiplexed signal $x_l(t)$ has the same bandwidth as the input signal $s_l(t)$. Additionally, according to the Eqs. (\ref{output at sampling points 1}) - (\ref{output at sampling points 5}) and (\ref{sampling theorem for a single channel}), except of the factor $\frac{1}{N}$ the sampling points are the same. Therefore, the following reconstruction formula can be concluded:
\begin{equation}\label{final equation}
\begin{aligned}
x_{l}(t)&=\left[\mathcal{F}_f^{-1}\left(\left[\mathcal{F}_t(s(t)\cdot\mathrm{sq}_{N,B}(t-l\Delta T))\right](f)\cdot\Pi_{B/N}\left(f\right)\right)\right](t)
\\
&=\Biggl[\mathcal{F}_f^{-1}\Biggl(\Biggl[\mathcal{F}_t\Biggl(\Biggl(\sum_{k=0}^{N-1} s_{k}(t)\cdot\mathrm{sq}_{N,B}(t-k\Delta T)\Biggr)
\cdot\mathrm{sq}_{N,B}(t-l\Delta T)\Biggr)\Biggr](f)\cdot\Pi_{B/N}\left(f\right)\Biggr)\Biggr](t)
\\
&=\frac{s_{l}(t)}{N}\ .
\end{aligned}
\end{equation}
Thus, the single channel $s_l$ of the multiplexed channels $s_1,...,s_{N-1}$ can be reconstructed without any loss of information.

\bibliography{reference}

\section*{Acknowledgements}
Arijit Misra and Thomas Schneider would like to gratefully acknowledge the financial support of the German Research Foundation (reference number: SCHN 716/15-2). 

\section*{Author contributions statement}
T.S. presented the basic idea of the agnostic sampling transceiver. S.P. and A.M. conceived the experiment,  A.M., S.P. and K.S. conducted the experiment, A.M. analysed the results. J.M. and T.S. presented the theory. All authors contributed in writing the manuscript. 

\section*{Additional information}
\textbf{Competing interests :} The authors declare no financial or non financial competing interests.
\end{document}